\documentclass[
superscriptaddress, 
aps,
pra,
twocolumn,
english]{revtex4-2}

\usepackage{graphicx}% Include figure files
\usepackage[dvipsnames]{xcolor}  %TODO
\usepackage{dcolumn}% Align table columns on decimal point
\usepackage{bm}% bold math
\usepackage{bbold}
\usepackage{amsmath}
\begin{document}

\preprint{APS/123-QED}

\title{Self-supervised learning for denoising quasiparticle interference data}% Force line breaks with \\
%\thanks{A footnote to the article title}%

\author{Ilse S. Kuijf}
\email{kuijf@lorentz.leidenuniv.nl}
\affiliation{Leiden Institute of Physics, Leiden University, The Netherlands}
\affiliation{$\langle aQa \rangle$ at Lorentz Institute and Leiden Institute of Advanced Computer Science, Leiden University, P.O. Box 9506, 2300 RA Leiden, The Netherlands}%Leiden University, P.O. Box 9506, 2300 RA Leiden, The Netherlands

\author{Willem O. Tromp}
\affiliation{Leiden Institute of Physics, Leiden University, The Netherlands}

\author{Tjerk Benschop}
\affiliation{Leiden Institute of Physics, Leiden University, The Netherlands}

\author{Niño Philip Ramones}
\affiliation{National Institute of Physics, University of the Philippines Diliman, Quezon City 1101, Philippines}

\author{Miguel Antonio Sulangi}
\affiliation{National Institute of Physics, University of the Philippines Diliman, Quezon City 1101, Philippines}

\author{Evert P.L. van Nieuwenburg}
\affiliation{Leiden Institute of Physics, Leiden University, The Netherlands}
\affiliation{$\langle aQa \rangle$ at Lorentz Institute and Leiden Institute of Advanced Computer Science, Leiden University, P.O. Box 9506, 2300 RA Leiden, The Netherlands}

\author{Milan P. Allan}
\affiliation{Leiden Institute of Physics, Leiden University, The Netherlands}
\affiliation{Faculty of Physics, Ludwig-Maximilians-University Munich, Munich 80799, Germany}
\affiliation{Center for Nano Science (CeNS), Ludwig-Maximilians-University Munich, Munich 80799, Germany}
\affiliation{Munich Center for Quantum Science and Technology (MCQST), Ludwig-Maximilians-University Munich, Munich 80799, Germany}

\date{\today}

\begin{abstract}
Tunneling spectroscopy is an important tool for the study of both real-space and momentum-space electronic structure of correlated electron systems. 
However, such measurements often yield noisy data. 
Machine learning provides techniques to reduce the noise in post-processing, but traditionally requires noiseless examples which are unavailable for scientific experiments. 
In this work we adapt the unsupervised Noise2Noise and self-supervised Noise2Self algorithms, which allow for denoising without clean examples, to denoise quasiparticle interference data. 
We first apply the techniques on simulated data, and demonstrate that we are able to reduce the noise while preserving finer details, all while outperforming more traditional denoising techniques. 
We then apply the Noise2Self technique to experimental data from an overdoped cuprate ((Pb,Bi)$_2$Sr$_2$CuO$_{6+\delta}$) sample.
Denoising enhances the clarity of quasiparticle interference patterns, and helps to obtain a precise extraction of electronic structure parameters. 
Self-supervised denoising is a promising tool for denoising quasiparticle interference data, facilitating deeper insights into the physics of complex materials.
\end{abstract}

%\keywords{Suggested keywords}%Use showkeys class option if keyword
                              %display desired
\maketitle

% Jan: Use "The Block Method: Writing Scientific Papers Without Tears"

\section{\label{sec:intro}Introduction} 

A common denominator of strongly correlated systems is the presence of emergent phenomena. 
The plethora of states of the copper-oxide superconductors is one of the most notorious examples of this~\cite{keimer2015quantum}. Several phases are not yet understood: strange metals, the pseudogap, and possibly most importantly, superconductivity.
%Much remains unknown about the mechanisms responsible for the complex high-temperature superconducting state. 

A tool for the investigation of strongly correlated systems is quasiparticle interference (QPI)~\cite{hoffman2002imaging,mcelroy2003relating,kohsaka2008cooper,fujita2014simultaneous,he2014fermi}, bridging real-space and momentum-space electronic structure~\cite{capriotti2003wave,wang2003quasiparticle,zhu2004power}. 
It does so with a high energy resolution for both occupied and unoccupied states, that is limited by thermal energy. 
However, QPI experiments are challenging, and the resulting data often suffers from significant noise levels, for reasons that we discuss below.

Developing noise reduction techniques is a long-standing yet continuously evolving field of research~\cite{goyal2020image}. 
Recently, unsupervised and self-supervised machine learning denoising techniques have advanced rapidly and risen as a tool to reduce experimental noise~\cite{batson2019noise2self, krull2019noise2void, hendriksen2020noise2inverse, laine2019high, hendriksen2021deep, ulyanov2018deep}. 
These type of machine learning denoising techniques have already been applied to fluorescence microscopy~\cite{batson2019noise2self, zhang2019poisson}, ARPES~\cite{kim2021deep, liu2023removing} and transmission electron microscopy data~\cite{krull2019noise2void, wang2020noise2atom}. 
Related are recent works focusing on the recovery of phase-sensitive information from full experimental QPI images, such as employing blind deconvolution~\cite{cheung2020dictionary} and multi-atom techniques~\cite{sharma2021multi}.

In this work, we apply self-supervised denoising techniques to both simulated and measured experimental QPI data. 
To do so, we adapt the Noise2Self~\cite{batson2019noise2self} algorithm to the QPI problem. 
We test the results against the ground-truth that is known for the simulated data, and address the quality of denoised experimental QPI data. 
We then use denoised QPI data from a cuprate superconductor, (Pb,Bi)$_2$Sr$_2$CuO$_{6+\delta}$ (BSCCO), to characterize the changes in momentum space electronic structure with increased overdoping.

\begin{figure*}
    \centering
    \includegraphics[width=1.0\textwidth]{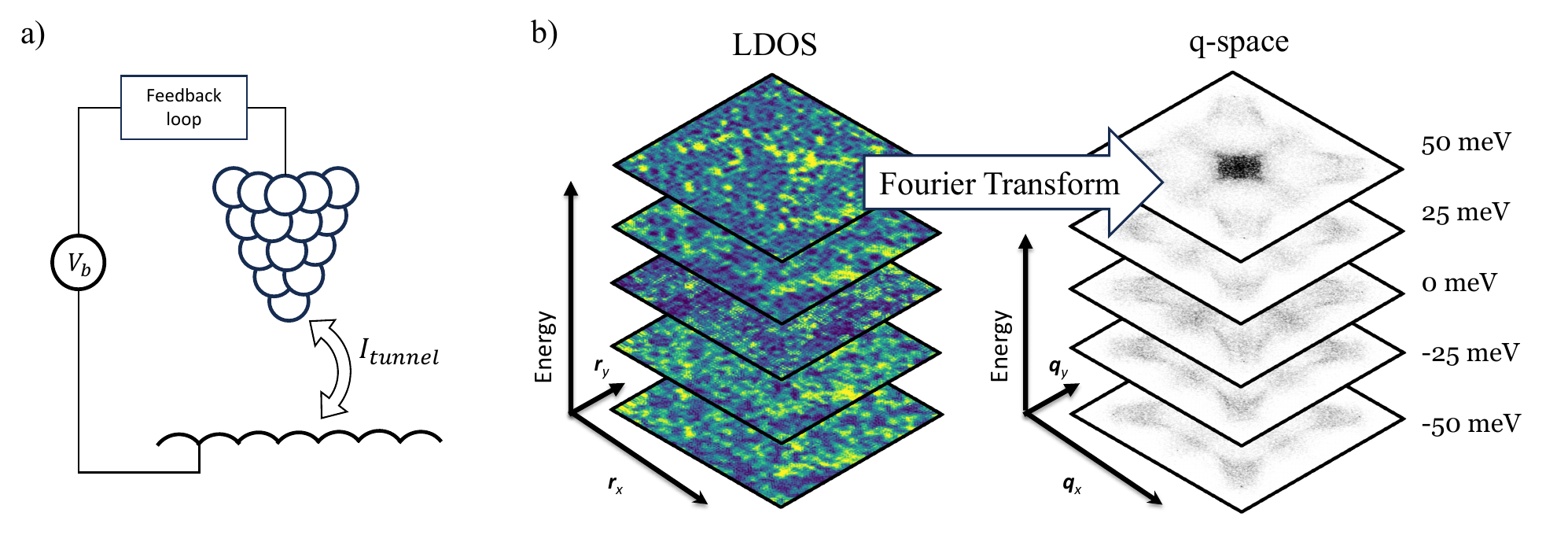}
    \caption{\textbf{Scanning tunneling spectroscopy.} (a) Schematic of the tip-sample system for scanning tunneling microscopy, showing the tunneling current $I_{tunnel}$ as result of the voltage bias $V_b$ between tip and sample. (b) Example of a 3-dimensional $dI/dV$ measurement, represented as images at different energy levels in both real- and $q$-space.}
    \label{fig:stm_setup}
\end{figure*}

\section{\label{sec:theory}Theoretical background}

\subsection{Quasiparticle interference}

Scanning tunneling spectroscopy is a powerful tool used to study the electronic properties of materials at the atomic scale.
In scanning tunneling spectroscopy, an atomically sharp metallic tip is brought into tunneling contact with the sample (Fig.~\ref{fig:stm_setup}a). 
A bias voltage between the tip and sample allows a tunneling current to flow from occupied states on one side to unoccupied states on the other side. 
The differential conductance $dI/dV$ can then be measured directly by adding a small modulation $\Delta V$ at a certain frequency, and using a lock-in amplifier to measure the resulting $\Delta I$ modulations at that frequency. 
This conductance measurement is directly proportional to the local density of states (LDOS) of the sample beneath the tip. 
At each measurement point, the tip position is stabilized and the bias voltage is swept while measuring the current. 
Sweeping the bias voltage allows a measurement of the LDOS at different energy levels, creating a three-dimensional map of the LDOS (Fig.~\ref{fig:stm_setup}b). 

While scanning tunneling spectroscopy is conventionally used to study the real-space electronic structure, it can also be used to measure the momentum-resolved structure.
This can be done through a technique called QPI, which is the result of elastic scattering of quasiparticles on defects and impurities in the crystal lattice. 
In elastic scattering, energy is conserved, which causes eigenstates of different $\vec{k}$ with the same energy $E(k)$ to become mixed. 
If a quasiparticle with initial $\vec{k_1}$ scatters to a state with $\vec{k_2}$, the interference between these states will cause an oscillation in the LDOS with a wavevector $\vec{q}=\vec{k_1}-\vec{k_2}$. 
This wavevector $\vec{q}$ can be measured by imaging the interference pattern in real space and taking a Fourier transform to reveal the $q$-space. 
The measured $q$-vectors can be related back to the band structure, in general only with some prior knowledge of said band structure. 
However, in the case of a single-band material as BSCCO, the $q$- and $k$-vectors are simply related as $\vec{q}=2\vec{k_f}$.

\subsection{Formulating the QPI noise problem}
The QPI signals typically show a considerable amount of noise, complicating the analysis of the data. 
To simplify the analysis, it is thus helpful to reduce the noise levels. 
This can be done using machine learning techniques, in which we train a model to remove noise from input data. 
It is important to understand the source and properties of the noise, as some machine learning algorithms require certain assumptions to hold. 

Suppose that a single scattering center creates an interference pattern $\rho_0(\vec{r},E)$, where $\vec{r}$ is the position and $E$ is the energy. 
For an experiment where multiple scatterers are distributed according to some potential $A(\vec{r}$), the resulting real-space image $\rho(\vec{r},E)$ is then:
\begin{equation}
    \rho(\vec{r},E)=\rho_0(\vec{r},E) \ast A(\vec{r})+\epsilon.
\end{equation}
In this equation $\ast$ is the convolution operator and $\epsilon$ is spatially uncorrelated experimental noise, such as thermal, electrical and mechanical noise. 
Experimental noise sources which have a correlated effect on groups of pixels, such as drift and tip imperfections, are ignored in this description. 
These types of noise are outside the scope of our denoising algorithm and have to be corrected for through different means, such as drift correction and good tip preparation. 

The QPI image $\tilde{\rho}(\vec{q},E)$ is obtained by Fourier transforming
\begin{equation}
    \tilde{\rho}(\vec{q},E) = \int d\vec{r}\;e^{-i\vec{q}\cdot\vec{r}}\rho(\vec{r},E) = \tilde{\rho_0}(\vec{q},E)\mathcal{P}(\vec{q})\;+\tilde{\epsilon},
    \label{eq:qpi_phase_noise}
\end{equation}
where $\tilde{\epsilon}$ is the Fourier transform of $\epsilon$. 
The Fourier transform of a single scatter, $\tilde{\rho}_0(\vec{q},E)$, is the clean QPI image which we hope to reconstruct. 
With respect to this clean image, the experimental QPI image contains some uncorrelated experimental noise $\tilde{\epsilon}$, and a complex valued phase factor $\mathcal{P}(\vec{q})$. 
In theory this phase factor is a multiplicative correlated noise, due to it being a Fourier transform of some potential map $A(\vec{r})$. 
However, for many aperiodically, non-uniformly distributed scatters, this convolutional noise resembles a random speckle noise (see Appendix~\ref{app:speckle_noise}), and we treat it as uncorrelated noise. 
This means that experimental QPI images contain both a multiplicative and an additive noise term, which are both approximated as spatially uncorrelated.

\subsection{Self-supervised denoising for QPI data}

\begin{figure*}
    \centering
    \includegraphics[width=1.0\textwidth]{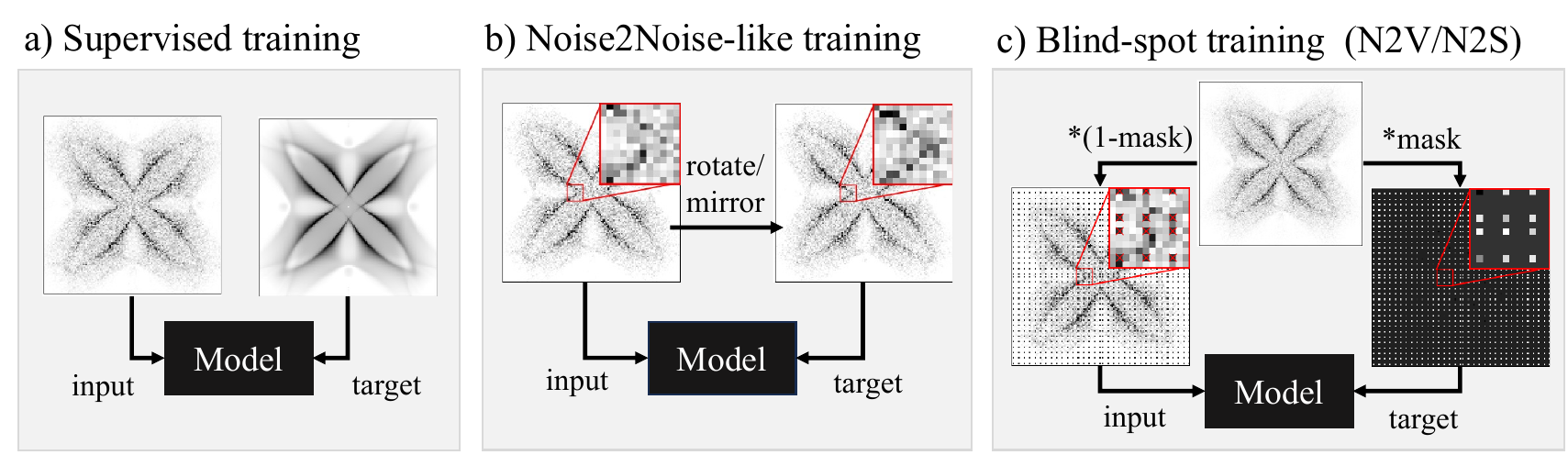}
    \caption{\textbf{Machine learning techniques for noise suppression.} (a) Supervised approach using clean images. (b) Noise2Noise-like self-supervised approach for QPI data, using geometric transformations. (c) Noise2Void/Noise2Self self-supervised approach using a special masking strategy.}
    \label{fig:ml_explanation}
\end{figure*}

Experimental noise can often be reduced by measuring over longer times or collecting more data to average over (for uncorrelated noise). 
Unfortunately the measurement time in our experiments is already pushed to its limit, as QPI requires scanning a large sample area at high spatial resolution.
Furthermore, more averages cannot get rid of the multiplicative noise from the phase term $\mathcal{P}(\vec{q})$ that arises from the Fourier transform. 
An alternative could be to smooth the image with a moving window, but this inevitably leads to broadening of the features in the signal, and can even lead to accidental removal of faint features. 
Hence, to suppress the noise while preserving the sharpness, we implement denoising through machine learning techniques.

The state of the art strategy for training a denoising model is to use a dataset consisting of pairs of noisy input images $\hat{x}_i$ and clean ground-truth target images $y_i$ (Fig.~\ref{fig:ml_explanation}a). 
With such a dataset we can train a parametrised model $f_\theta$ in a supervised learning setup, where the general task is to find the optimal model $f^*_\theta$:
\begin{equation}
    f^*_\theta = \text{argmin}_\theta\, \sum_i L\big(f_\theta(\hat{x}_i), y_i\big).
    \label{eq:the_learning_problem}
\end{equation}
In this equation, the sum runs over all samples in the dataset and $L$ is the loss function, which is typically taken to be the pixel-wise mean-squared error.
The model $f_\theta$ is typically a convolutional neural network, because the convolutional layers are particularly effective at extracting spatial structures from the data. 

A major challenge in applying supervised learning to denoise scientific data is that the ground-truth images $y_i$ are often unavailable. 
This could be resolved with a simulation-based approach~\cite{joucken2022denoising, mohan2022deep}, but such an approach requires assumptions about the noise model and physics involved, and suitable simulations are not always available or feasible.
Due to this lack of ground-truth images for the supervised approach, unsupervised learning techniques have been developed that don't rely on ground-truth images.
These techniques can train a model to denoise images using only the noisy versions of the data~\cite{krull2019noise2void, lehtinen2018noise2noise, batson2019noise2self}. 

Noise2Noise~\cite{lehtinen2018noise2noise} is such a technique, which can reconstruct a cleaner image from multiple independent instances of a noisy image.
The training proceeds similarly to Eq.~\ref{eq:the_learning_problem}, but with the clean targets $y_i$ replaced by corrupted versions $\hat{y}_i$.
Not all experimental setups allow for an easy acquisition of multiple independent instances of the noise, because e.g. the experiment might be unstable, single-shot, or have a very long measurement time.
In our case we can create multiple instances of the noise because of the inherent symmetry of the QPI signal, either in momentum-space or by dividing the real-space image into multiple subsets and creating a noisy image for each (similar to Noise2Inverse~\cite{hendriksen2020noise2inverse}).
In the following, the Noise2Noise results we present are obtained by rotating and mirroring the momentum-space images (Fig.~\ref{fig:ml_explanation}b), yielding four `different' versions per image.
Note that although the resulting images are not independent, we expect that the local differences are sufficient for the Noise2Noise approach to be effective.

Additionally, we also investigate self-supervised techniques designed to deal with the scenario where multiple independent instances of the noise are unavailable. 
Both Noise2Void~\cite{krull2019noise2void} and Noise2Self~\cite{batson2019noise2self} are such techniques, in which the same noisy image is used as both the input and the target when training the model.
Self-supervision works by using a blind-spot masking scheme (Fig.~\ref{fig:ml_explanation}c), where a subset of pixels in the input image is masked, and the complement is given to the model as input. 
Denoting the set of masked pixels by $J$, and the complement (i.e. all other pixels) by $J^C$, the masked image is constructed as $\mathbb{1}_{J^C}\hat{x}_i$, and the pixels in the mask itself as $\mathbb{1}_J\hat{x}_i$. 
Here $\mathbb{1}_J$ represents the indicator function, which, when applied element-wise to an image, preserves pixels in $J$ and sets others to zero.
The model is tasked with predicting the masked pixels from the masked image, for several different masks.
This means it needs to learn to predict the masked pixels by using the information of the neighboring pixels.
The learning problem for the self-supervised case is then:
\begin{equation}
    f^*_\theta = \text{argmin}_\theta\, \sum_i \sum_J ||\mathbb{1}_Jf_\theta(\mathbb{1}_{J^C}\hat{x}_i) - \mathbb{1}_J\hat{x}_i ||^2
\end{equation}
In this equation the sum over $J$ represents the different masks that are averaged over, such that each pixel gets masked once.
To create fair masks for QPI data, we employ a rotationally symmetric grid to create the subsets, to take into account the inherent rotational invariance of the data caused by the Fourier transform. 

The blind-spot techniques are based on the assumption that the noise is mean-zero and element-wise statistically independent. 
For QPI there are no consequences if the mean-zero assumption does not hold, as this would result in a global offset in the denoised image, and we are only interested in correlations in the image, not the quantitative values. 
The element-wise statistical independence assumption is needed because the algorithms work by extracting correlations in the image. 
If the noise is correlated, it would extract the noise as a signal, and hence fail to remove the noise. 
For QPI data, this means that noise sources such as drift and tip imperfections cannot be corrected for using these techniques.
These algorithms are fully model-agnostic and make no further assumptions about the noise model or the physics underlying the images. In the following, we use an adapted version of the Noise2Self technique.

\section{Results}

\subsection{Denoising simulated data}

\begin{figure*}[!ht]
    \centering
    \includegraphics[width=\textwidth]{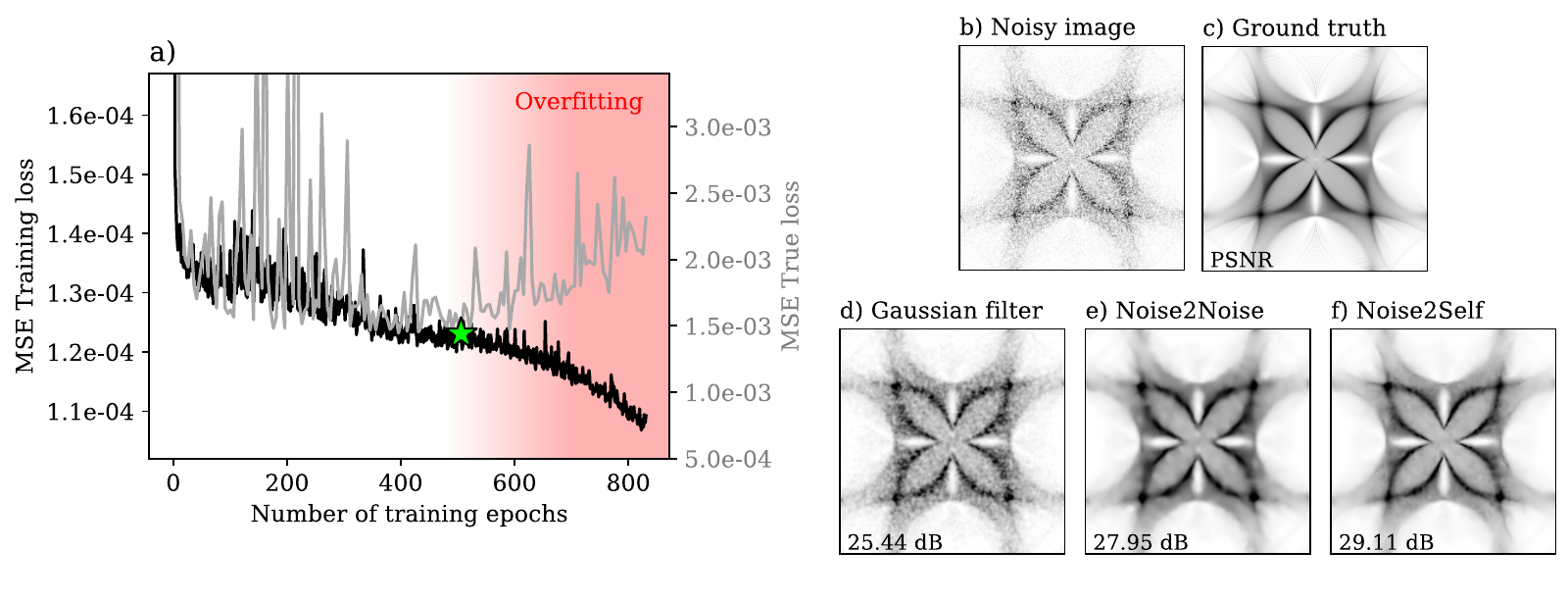}  %simulation_examples_sym.pdf
    \caption{\textbf{Denoising results for the simulated dataset.} he mean-squared errors for the Noise2Self method between the denoised outputs and noisy inputs (black) and between the denoised outputs and the ground-truth images (gray). The green star indicates the point of optimal training (lowest true MSE). (b) Example of a noisy image. (c) Ground truth of the sample image. (d-f) The denoised results for the Gaussian filtering, Noise2Noise, and Noise2Self methods, respectively. The bottom left corner of each denoised image shows the PSNR of that image with respect to the ground truth. }
    \label{fig:sim_results}
\end{figure*}

Given the absence of a known ground truth for experimental data, verifying the accuracy of denoised results is a challenge. 
This lack of a reliable objective method for comparing results complicates tuning hyperparameters to optimize the denoised results. 
We address these problems by tuning the denoising algorithm on simulations of similar data, where the quality of the results can readily be tested. 
In this section, we discuss denoising results for simulated QPI data, and we determine the best strategy for denoising QPI data.

We perform simulations of noisy QPI data by computing the local density of states using real-space Green’s functions defined on two-dimensional lattices, with a smooth many-impurity disorder potential~\cite{sulangi2017revisiting}. 
The corresponding clean data is simulated using a single smooth scatterer. 
The simulations are based on a tight-binding Hamiltonian for a $d$-wave superconductor on a square lattice:
\begin{equation}
\begin{aligned}
    H &= \mu \sum_{i\sigma} c^\dagger_{i\sigma}c_{i\sigma} \\
    &\phantom{=} + \sum_{\langle i, j \rangle} \Big[ - \sum_{\sigma \in \{\downarrow, \uparrow\}} t_{ij}c^\dagger_{i\sigma}c_{j\sigma} + \Delta_{ij}c^\dagger_{i\uparrow}c^\dagger_{j\downarrow} + \Delta^*_{ij}c_{i\uparrow}c_{j\downarrow}\Big]
\end{aligned}
\end{equation}
For $t_{ij}$, nearest-neighbor and next-nearest-neighbor hoppings $t_1$ and $t_2$ are included. 
The $d$-wave order parameter $\Delta_{ij}$ has the form $\Delta_{ij}=\pm\Delta_0$ (with positive and negative values for order parameters along the $x$- and $y$-directions, respectively) to incorporate $d$-wave pairing. 
A chemical potential $\mu$ is also included. 
A dataset consisting of 480 pairs of different images was used to test the denoising algorithms. 
The exact parameters used to simulate the data can be found in Appendix~\ref{app:sim_params}.

We determine the best denoising strategy by comparing several techniques. 
First, we consider the traditional technique of applying a Gaussian filter, where we cut out the center pixel to allow optimizing the filter size with the self-supervised loss.
We test sigmas from 0.2 to 1.5 pixels in 0.1 increments, and find the median optimal sigma for the entire simulated dataset to be 0.9 pixels.
For the machine learning techniques, a Noise2Noise-like approach and adapted version of Noise2Self are used. 
The Noise2Noise approach uses rotated and mirrored versions of the noisy image as `different' noisy versions of the same image. 
The Noise2Self technique is adapted to have rotationally symmetric masks, where masked pixels are replaced by an average of the neighboring eight pixels. 

For both the Noise2Noise and Noise2Self approaches, we use a DnCNN architecture with 14 convolutional layers with 64 features each. The architecture follows the one of the DnCNN used by Batson \& Royer in the Noise2Self paper\cite{batson2019noise2self}.
We apply min-max scaling to scale the input data between 0 and 1.
The dataset is not split into a train and test set, as the goal is to optimize performance on this specific data, rather than generalize to images outside the dataset. Due to the limited amount of images available, we want to utilize all available data for training.
The models are trained using the Adam optimizer with a learning rate of 0.005 and a batch size of 32.
We monitor the ground-truth mean-squared error (MSE) during training, and use it to determine the optimal training point.

Figure \ref{fig:sim_results}a shows that for the Noise2Self method the optimal training point is reached after 505 epochs, after which the performance deteriorates. 
Visual inspection of the model output during training shows that when the model is trained for too long, it starts to overfit on the noise. 
For an example of this overfitting, see Appendix~\ref{app:overfitting_noise2self}.
A reason for the overfitting of the Noise2Self method could be that the phase factor $\mathcal{P}(\vec{q})$, which we assumed to be spatially uncorrelated noise, contains some spatial correlations which cause the model to extract this noise as signal after longer training times.
The Noise2Noise method does not suffer from this overfitting, which could be explained by this method making no assumptions about the structure of the noise.
Figure~\ref{fig:sim_results}b shows an example of a simulated noisy QPI signal, Figure~\ref{fig:sim_results}c shows the corresponding ground-truth signal, and Figures~\ref{fig:sim_results}d-f show denoised versions for the different techniques. 
The peak signal-to-noise ratios (PSNRs), shown in the corners of the figures, show that Noise2Self achieves the best denoising result for this sample. 
More examples can be found in Appendix~\ref{app:extra_simulated_figs}. 

Over the entire simulated dataset, the Gaussian filter achieves an average PSNR of 27.8~dB, the Noise2Noise technique 29.7~dB, and Noise2Self 30.9~dB. 
From this, we can conclude that self-supervised denoising using a blind-spot method like Noise2Self outperforms traditional data denoising using Gaussian filtering. 
Visual inspection of the denoised images shows that this method gets results which closely resemble the ground-truth images, and preserves and reconstructs finer features in the data well.

\subsection{Experimental results}

Having established the Noise2Self method as the best strategy for denoising QPI data using simulated data, we now move on and apply it to experimental data. 
For this, we chose the overdoped cuprate high-temperature superconductor (Pb,Bi)$_2$Sr$_2$CuO$_{6+\delta}$ (BSCCO). 
These materials host a number of open questions relevant to condensed matter physics, and in particular exhibit unusual quasiparticle interference ~\cite{keimer2015quantum,sulangi2017revisiting,Sulangi2018,Sulangi2018b,Zheng2017,SHEN200814}. 
We apply our denoising algorithm to a number of datasets of this material and extract important parameters including electron density and dispersion by fitting the resulting denoised data.

For the measurements, the crystals were cleaved at 4~K and in ultra-high vacuum ($<$ 1e-10~mbar), and transferred directly into our home-built scanning tunneling microscope (T = 4.2~K) ~\cite{battisti2018definition}; for more details, see Ref. ~\cite{Tromp2023}. 
We measure several samples in the overdoped (OD) region of the phase diagram, with superconducting transition temperatures ($T_c$) of 23~K, 12~K and 3~K respectively.

We measure large-scale conductance maps ($g(\vec{r},E)=dI/dV$) using the conventional lock-in technique. We correct our data for drift using the Lawler-Fujita algorithm \cite{lawler2010intra, fujita2014direct}.
Before taking the Fourier transform, the conductance maps are often normalized to reduce set-up effects. To study particle-hole symmetric features, the ratio $Z(\vec{r},E) = g(\vec{r},+E)/g(\vec{r},-E)$ is often taken. To study bands, $N(\vec{r},E) = g(\vec{r},E)/\big(I(\vec{r},E)/V\big)$ is also used (where $I(\vec{r},E)$ is the current map). For discussions, see e.g.\ Refs.~\cite{FEENSTRA1987, battisti2020direct, fujita2014spectroscopic, yazdani2016spectroscopic}.
In Appendix~\ref{app:comparison_gNZ}, we make a brief comparison between $g(\vec{r},E)$, $N(\vec{r},E)$ and $Z(\vec{r},E)$. In this study, we use $N(\vec{r},E)$. 

\begin{figure*}
    \centering
    \includegraphics[width=0.85\textwidth]{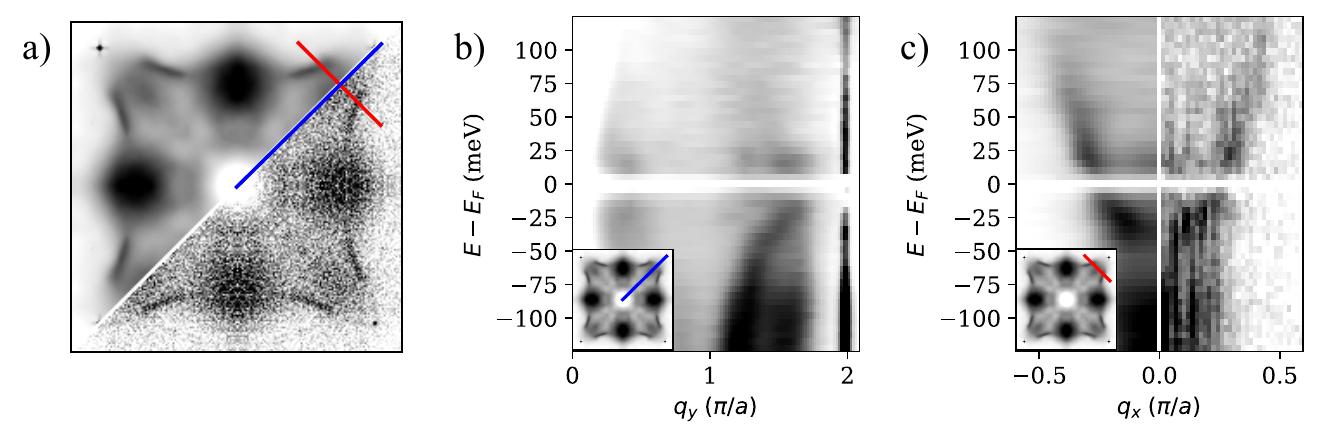}
    \caption{\textbf{Dispersions along different cuts for the OD3K sample.} (a) Fermi layer indicating where the cuts are taken, showing half the symmetrized denoised result (top left) and half the symmetrized original noisy image (bottom right). (b) Cut of the denoised data from the (0,0) to (0, $2\pi/a$) point. Inset shows cut location. (c) Antinodal cut, symmetrized, showing half denoised and half original noisy image. Inset shows cut location.}
    \label{fig:clean_linecuts}
\end{figure*}

We then calculate the Fourier transform of each layer $N(\vec{q},E)$, revealing the QPI patterns commonly observed in overdoped BSCCO samples~\cite{he2014fermi}. Note that because of the Pb substitution, the supermodulation typical for BSCCO is absent, facilitating an easier interpretation of the measured QPI patterns. 
In Fig~\ref{fig:clean_linecuts}a (bottom right half), $N(\vec{q},E)$ is shown for the OD3K sample (meaning overdoped BSCCO with $T_c$ = 3~K). 

To prepare the data for the denoising routine, the low-$q$ core is suppressed by applying an inverse Gaussian filter in $q$-space. The same filter width is used for the entire dataset.
Extreme pixel values which might otherwise negatively influence the learning are capped by identifying the 99th percentile value, and capping any values above this threshold to the 99th percentile.
To treat the data as images, we take the absolute value after Fourier transforming. 
To improve model performance, the measurements of the different samples (OD3K, OD12K, OD23K) are combined into a single training dataset of 183 different images. 
In order to combine the different measurements, they are rescaled to the same real-space resolution and cut to the same size of 192x192 pixels.
We further expand the dataset by using geometrical data augmentation techniques (rotation, mirroring), resulting in a total dataset of size 732x192x192. 
A DnCNN (Denoising Convolutional Neural Network)~\cite{zhang2017beyond} consisting of 14 layers is trained for 75 epochs. 
The denoised result is shown in the upper-left corner of Fig~\ref{fig:clean_linecuts}a. 
In contrast to the denoised simulated datasets, we can no longer compare with the ground-truth images and a comparison is made instead with the raw experimental data and with low-pass filtered data (see Appendix~\ref{app:comparison_N2S_gaus}). 
It is apparent that the QPI pattern appears to be sharper, and with less noise after denoising. 

We will now discuss some prominent features that are visible. 
For details on quasiparticle interference data, interpretation and for more the nomenclature see e.g. ~\cite{he2014fermi,webb2019density}. 
Starting in the corner of the Brillouin zone (Fig.~\ref{fig:clean_linecuts}a), we observe a triplet feature which is typically associated with the large Fermi surface present in the overdoped regime of the cuprates. 
It was shown previously that this QPI feature mainly originates from scattering between antinodal points on the Fermi surface and is hence referred to as antinodal QPI~\cite{he2014fermi}. 
Interestingly, considering this antinodal QPI, our data seems to deviate from observations in this work in two distinct ways: 1) Our antinodal QPI does not appear to extend beyond the antiferromagnetic Brillouin zone, and instead fades out at the border. This seems to be the case for all of our datasets. However, it is important to point out that these effects can also depend on set-up effect and normalization method, and that at higher doping, when most of the Fermi surface is inside the antiferromagnetic Brillouin zone, the distinction becomes more challenging. 
2) The observed triplet feature appears to be dispersive, contrary to what was reported previously. 
In Fig.~\ref{fig:clean_linecuts}b, we show a line cut along the (0,2$\pi$/a) direction of the QPI data on the OD3K sample (blue line in Fig.~\ref{fig:clean_linecuts}a). 
The dispersive behavior is especially clear at energies further away from the Fermi level, in the occupied states. 
The QPI for the unoccupied states are not visible; these should appear beyond the Bragg peaks, but as noted earlier the signal vanishes for $q$-vectors beyond the Bragg peaks.

We also note the weakly dispersing signal in the nodal region, which is less sharp. The reason for that is that it originates from mixing between antinodal and nodal scattering and nodal-nodal scattering, which leads to a much broader area with signal. 

\begin{figure*}
    \centering
    \includegraphics[width=1.0\textwidth]{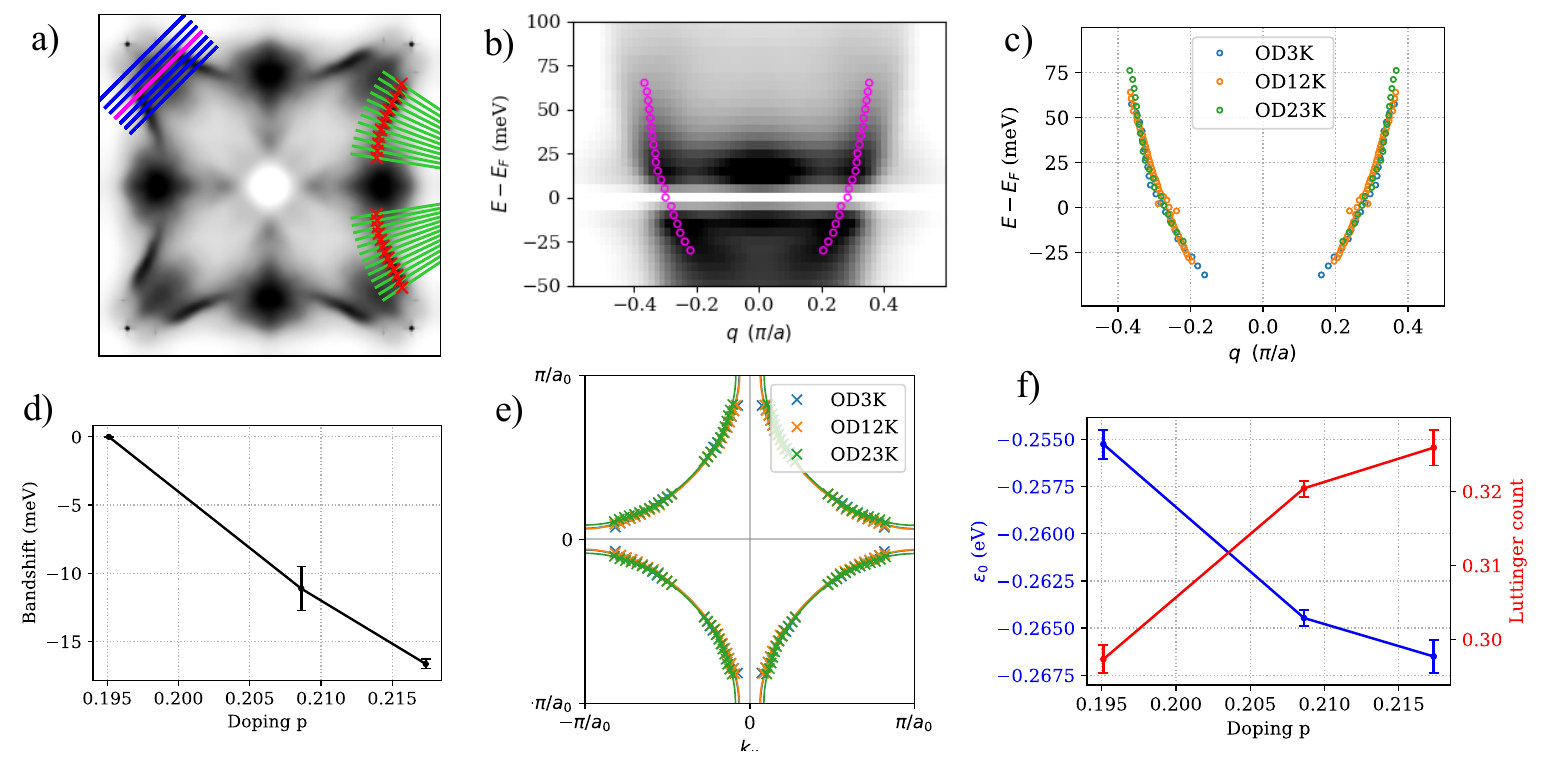}
    \caption{\textbf{Fits extracted from the denoised data.} 
    (a) Overview of the Fermi level of the OD23K sample, indicating the cuts used to extract the Fermi surfaces (green) and band shifts (blue/purple). 
    (b) Antinodal cut of the OD23K sample, indicated by the purple line in (a). Purple circles indicate the fitted peaks of the momentum distributions. 
    (c) Result of vertically shifting the extracted bands to minimize the distances relative to the OD23K band. The points shown are from the cut indicated in purple in (a). 
    (d) The averaged fitted relative band shifts for all blue/purple antinodal cuts shown in (a), as a function of hole-doping.
    (e) Fermi surfaces determined by transforming peaks extracted from the green radial cuts in (a) to $k$-space. The solid lines show tight-binding fits.
    (f) The $\epsilon_0$ parameter of the tight-binding fits shown in (e), and the Luttinger count calculated using the area of the fitted Fermi surface, as a function of hole-doping.}
    \label{fig:fits}
\end{figure*}

Next, we discuss the antinodal QPI in more detail. In Fig.~\ref{fig:clean_linecuts}c, we show a line cut orthogonal to the antinodal direction (red line in Fig.~\ref{fig:clean_linecuts}a). 
We can compare these cuts between datasets and extract the rigid band shift due to doping.
We do this by extracting the QPI peak positions for different antinodal cuts, indicated in Figure~\ref{fig:fits}a in blue/purple. 
For each sample, the extracted peak positions (Fig.~\ref{fig:fits}b) are compared to the cut obtained from the OD23K sample, minimizing the energy shift required to overlap the datapoints with those of the OD23K sample through a $\chi^2$-fit (Fig.~\ref{fig:fits}c). 
The fitted energy shifts are averaged over the different cuts shown in Figure~\ref{fig:fits}a, and the results are shown in Fig.~\ref{fig:fits}d, where we clearly see the bands shift with doping.
The doping levels shown in Figure~\ref{fig:fits} were calculated using the Ando formula~\cite{ando2000carrier}.

Next, we trace the Fermi surface for a number of energy layers around and including the superconducting gap. For each of the green cuts indicated in Figure~\ref{fig:fits}a, the peak positions $\vec{q}=2\vec{k}_F$ are fitted. The results of these fits are shown in Figure~\ref{fig:fits}e, where the vectors are rotated and divided by a factor 2 to obtain the corresponding $k$-vector. Figure~\ref{fig:fits}e also shows tight-binding fits to the extracted Fermi surfaces, where all parameters except the chemical potential $\epsilon_0$ were fixed, following~\cite{he2014fermi}. The three Fermi surfaces and their fits are nearly identical, particularly for the OD3K and OD12K samples. This is because the band shift at these doping levels is minimal, especially in the nodal region. The fitted values for $\epsilon_0$ are shown in Figure~\ref{fig:fits}f, together with the Luttinger counts that can be extracted from the area of the fitted Fermi surfaces. This, together with the observed rigid band shift, allows to measure doping and $T_c$ independently, and not, as it is often done, extract the latter from the former via the Ando or Presland formulas~\cite{presland1991general}. This has been done previously using data from photoemission experiments~\cite{drozdov2018phase}, and scanning tunneling microscopy~\cite{Tromp2023}.
Importantly, we find values for the Luttinger count, which are systematically higher than universal formulae would suggest based on $T_c$. Our values agree with ARPES results for BSCCO~\cite{berben2022superconducting, kondo2004hole}. These findings suggest that defining the doping axis with hole counts instead of estimated values is important.

\section{Conclusions}

In summary, we have applied Noise2Self denoising to QPI data for the first time. 
We describe in detail the results, both using simulated data where the ground truth is known, and with our experimental data. 
For the simulated datasets, we show that the denoised data are much closer to the ground truth than either raw data or low-pass filtered data. 
For the experimental data, such a comparison and thus a neutral way of testing the improvement is obviously not possible because of the lack of ground truth images. 
This connects to a general challenge with ML for data analysis in quantum materials research: it is often a black box for which interpretability is infeasible. 
Our approach in this paper is to test the method with simulated data that has a similar noise source. 
However, regarding experimental data, we show that the Noise2Self denoising respects the finer features of the data and produces a sharper image compared to, for example,  low-pass filtering of an image which is often done. 
A demonstration of this is for example shown in Appendix~\ref{app:comparison_N2S_gaus}, where we see that the Bragg peaks after denoising remain nicely sharp, whereas after Gaussian smearing, these become very blurred. 
We therefore think our technique could be beneficial for analyzing QPI data in any material.

\begin{acknowledgments}
This work was supported by the Dutch National Growth Fund (NGF), as part of the Quantum Delta NL programme.
\end{acknowledgments}

\appendix
\section{Examples of speckle noise pattern}\label{app:speckle_noise}
\begin{figure*}[!ht] %Note to self: figure code in original_code_miguel
    \centering
    \includegraphics[width=1.0\textwidth]{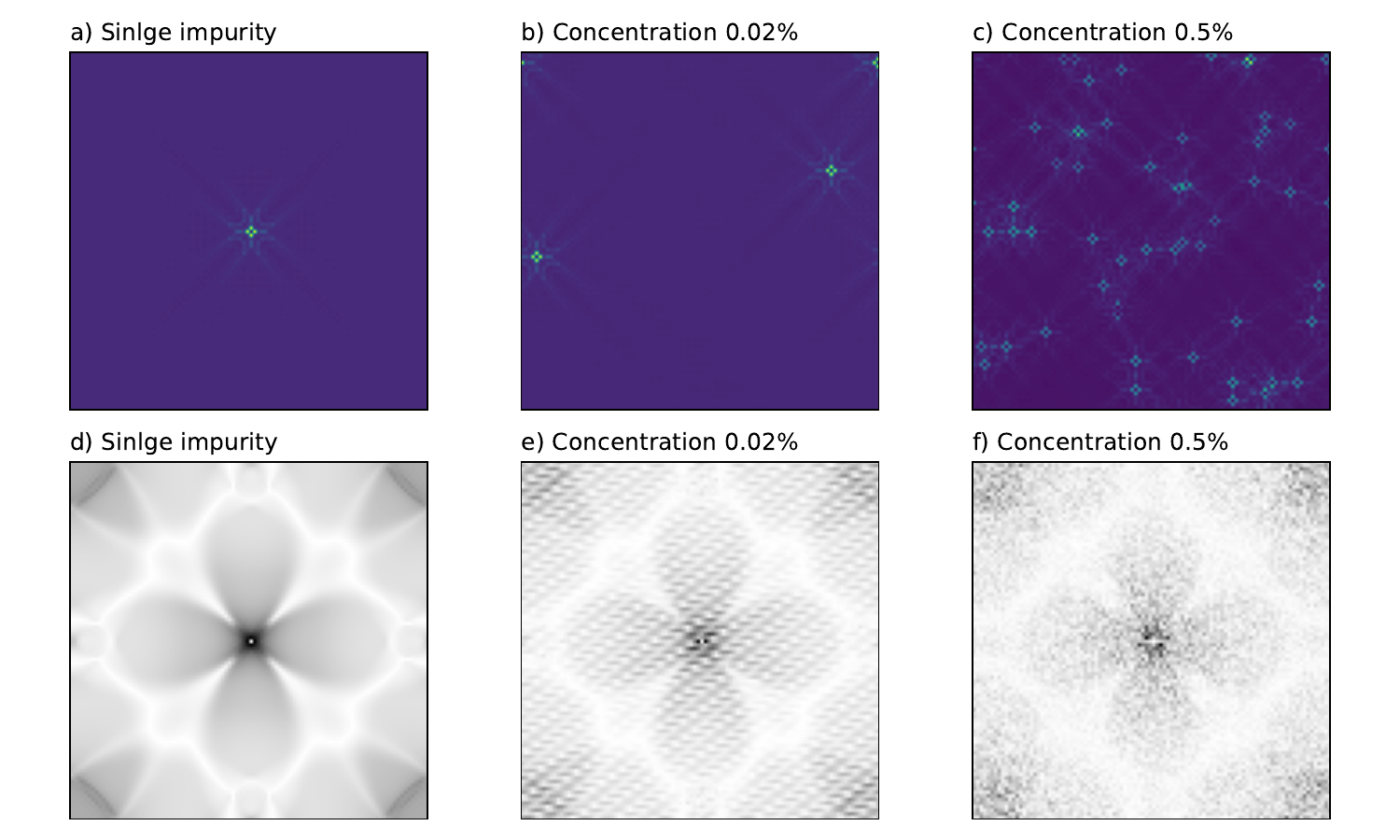}
    \caption{\textbf{Simulated examples of speckle noise pattern.} (a-c) LDOS images and (e-f) Fourier transformed images of a single impurity, a 0.02\% random impurity concentration, and a 0.5\% random impurity concentration, respectively.}
    \label{fig:sim_noise_illustration}
\end{figure*}

Experimental QPI images contain a speckle noise pattern in the momentum-space image, due to the presence of multiple complex scatterers which cause wave interference in the real-space image. This can be best illustrated with a simulation of pointlike scatterers, as is shown in Fig~\ref{fig:sim_noise_illustration}. Note that for cuprates, pointlike scatterers don't fully explain the experimental features observed~\cite{sulangi2017revisiting}. 

It can be seen that with many scatterers present in the field of view, the speckle noise, which is present in the signal due to the phase factor as shown in the main text equation~2, starts to resemble random noise. Experimental images will have a combination of this speckle noise and other noise sources such as thermal, electrical and mechanical noise. Self-supervised denoising can denoise all noise at the same time, as long as the noise is pixel-wise independent.

\section{Parameters used for the simulated dataset}\label{app:sim_params}
To achieve a high variety in images for the simulated dataset, we varied the different simulation parameters. The dataset has a total of 480 clean-noisy pairs of simulated images of size 192x192, which can be divided into 8 sets of 60 images. The tight-binding and disorder parameters are varied over the 8 sets (see Table.~\ref{tab:sim_params}). Within one set of 60 images, only the simulated energy level varies, while the model parameters and disorder potential are kept constant. This means the total dataset can be seen as a combination of 8 spectroscopy experiments of different random superconducting samples. The 60 simulated energy levels per set are linearly spaced from $-4\Delta$ to $4\Delta$.

For the disorder potential, we use an impurity concentration of 0.10, and the smooth disorder potential of the following Gaussian form:
\begin{equation}
V(\mathbf{r}) = \sum_i \frac{V_{imp}}{2\pi\sigma^2} e^{-\frac{1}{2\sigma^2}[(x - x_i)^2 + (y - y_i)^2]}
\end{equation}

\begin{table}[h]
    \centering
    \caption{Parameter values used to generate the simulated dataset.}
    \begin{tabular}{c|ccccc}
        Set & $t_1$ & $t_2$ &  $\Delta$&  $V_{imp}$& $\sigma$\\\hline
        1 & 1 &  -0.4&  0.08&  -0.1& 0.6\\
        2 & 1 &  -0.2&  0.07&  -1.0& 0.7\\
        3 & 1 &  -0.3&  0.08&  -1.0& 0.8\\
        4 & 1 &  -0.4&  0.07&  -1.0& 1.0\\
        5 & 1 &  -0.2&  0.08&  -0.5& 1.0\\
        6 & 1 &  -0.3&  0.08&  -0.5& 1.0\\
        7 & 1 &  -0.4&  0.07&  -0.5& 0.6\\
        8 & 1 &  -0.4&  0.08&  -0.5& 0.7\\
    \end{tabular}
    \label{tab:sim_params}
\end{table}

\section{Extra examples of different results simulated dataset}\label{app:extra_simulated_figs}
\begin{figure*}[!ht]
    \centering
    \includegraphics[width=1.0\textwidth]{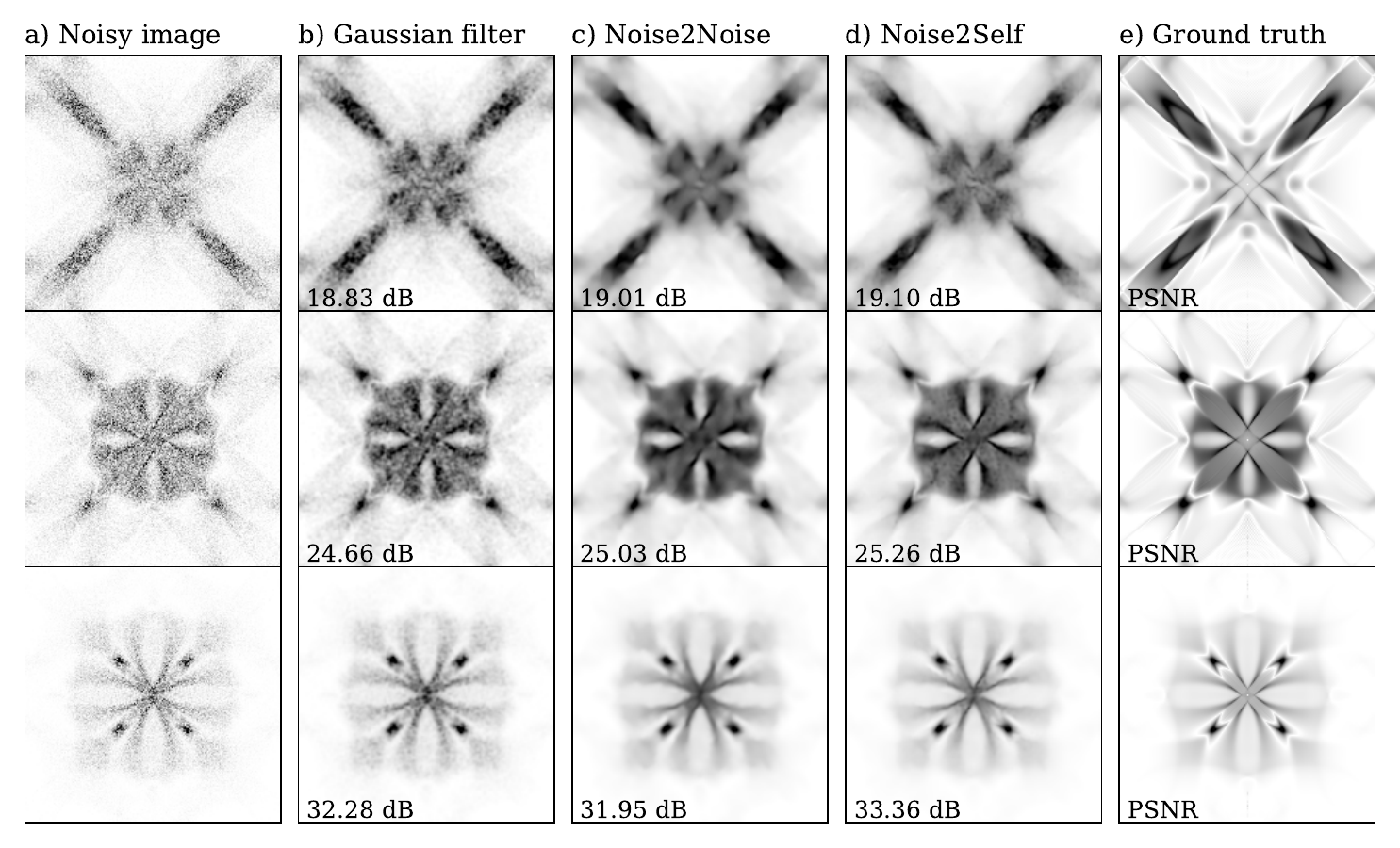}
    \caption{\textbf{Different denoising results for a sample from the simulated dataset.} Figure (a) shows the noisy image, (b-d) the denoised results for Gaussian filtering, Noise2Noise, and Noise2Self, and (e) shows the ground-truth image. The bottom left corner of each denoised image shows the PSNR of that image with respect to the ground-truth.}
    \label{fig:sim_examples_ext}
\end{figure*}

In the main text we showed an example image to visually asses the denoising results for the simulated dataset. Figure~\ref{fig:sim_examples_ext} shows additional examples. It can be seen that the same conclusions hold throughout the different examples. The images in the simulated dataset have a great diversity, capturing both fuzzy and sharp features. This shows that this method can work on many different types of QPI images.

\section{Overfitting of the Noise2Self method}\label{app:overfitting_noise2self}
\begin{figure*}[!ht]
    \centering
    \includegraphics[width=1.0\textwidth]{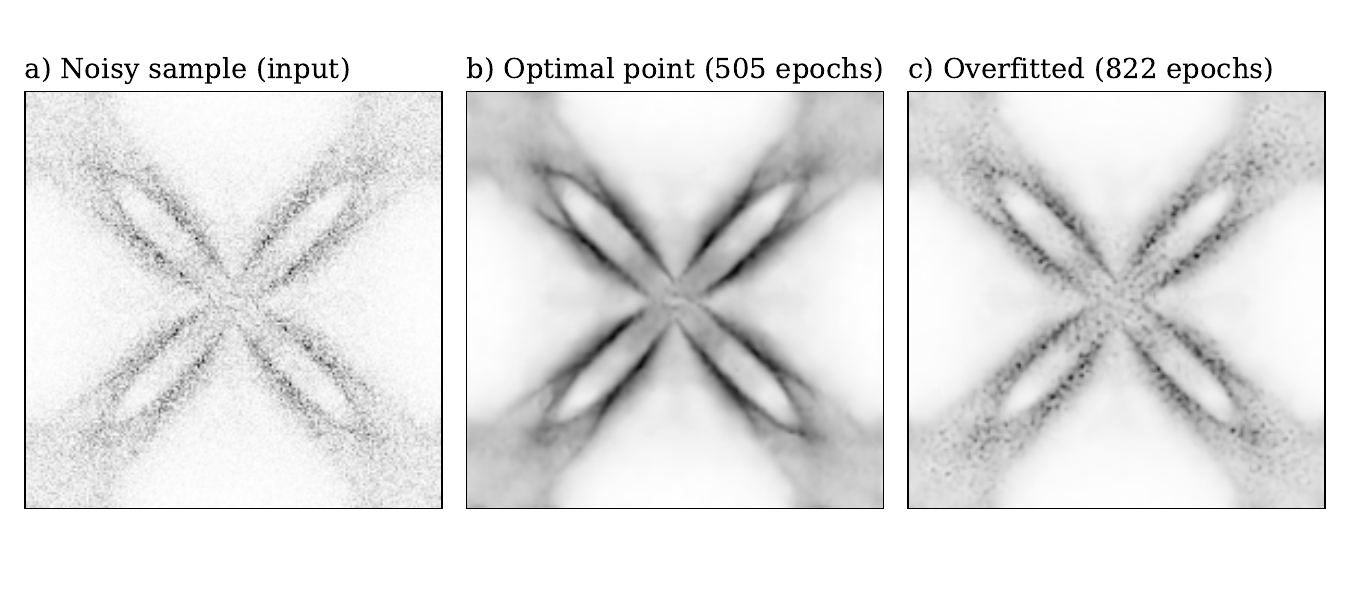}
    \caption{\textbf{Simulated image noisy, clean, and overfitted.} (a) The noisy input image. (b) The denoised output at the point of optimal denoising according to the ground-truth MSE. (c) The denoised output at the end of training, far into the overfitting regime.}
    \label{fig:sim_overfitted_example}

    \includegraphics[width=1.0\textwidth]{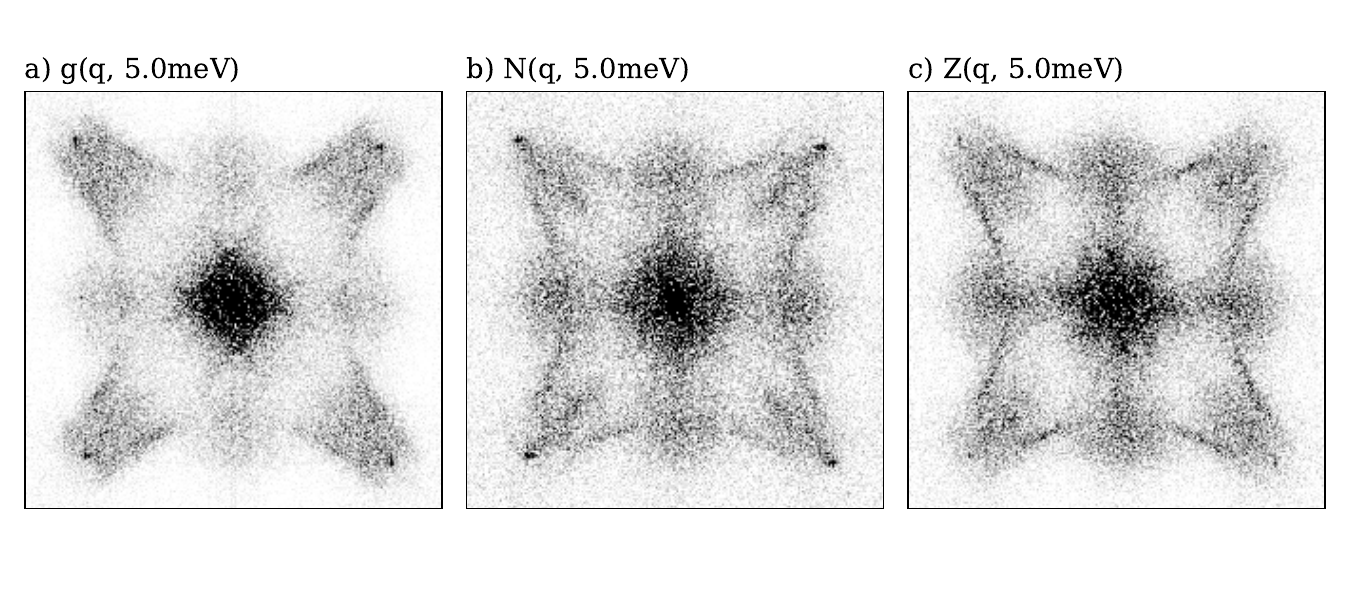}
    \caption{\textbf{Comparison of QPI patterns between $g$, $N$ and $Z$ maps, for the OD23K sample} (a) Fourier transform of the original conductance map $g$. (b) Fourier transform of the current-normalised map $N$. (c) Fourier transform of the ratio map $Z$.}
    \label{fig:gNZ_comparison}
\end{figure*}

Theoretically, the point where Noise2Self reaches a minimum in the self-supervised loss should be the point of optimal denoising. However, upon training several neural networks we obeserved that the self-supervised loss does not reach a minimum, and instead the network's outputs start to appear overfitted. Overfitting is a phenomenon which happens in machine learning when a network is trained for too long, past the optimal training point. In this case, overfitting means that the outputs start to appear noisy again (Fig.~\ref{fig:sim_overfitted_example}). 

The overfitting happens in both simulated and experimental data, which indicates it could either be a result of the denoising method or of the structure of QPI data. The training curves and outputs for the Noise2Noise-like technique showed no overfitting. In the denoising of experimental QPI data, where the ground-truth curve can not be calculated, the point of optimal denoising can be determined with a combination of visual inspection and inspection of the slope change in the loss curve.

\section{Comparison between g(r,E), N(r,E) and Z(r,E)}\label{app:comparison_gNZ}

Normalisation of the data plays an important role in mitigating the setup effect~\cite{FEENSTRA1987, macdonald2016dispersing}, which is an effect that causes artifacts due to a 'leak' of the signal at the setup bias into signals at other bias levels. For QPI measurements, this results in a constant-in-$\vec{q}$ artifact, of which the position and shape depend on the visible q-vector at the setup bias. 

To enhance the sharpness of the QPI features and mitigate the setup effect, the conductance measurements $g(\vec{r},eV) = dI/dV(\vec{r},eV)$ can be normalized with their corresponding current layers, by taking $N(\vec{r},eV)$ (Eq.~\ref{eq:norm_N}).

\begin{equation}
    N(\vec{r},eV) = dI/dV(\vec{r},eV)/(I(\vec{r},eV)/V)
\label{eq:norm_N}
\end{equation}

However, doing so results in a loss of definition close to the Bragg peaks. A comparison between $g$ and $N$ can be seen in Figure~\ref{fig:gNZ_comparison}a-b. Another method that can be used to enhance the QPI and cancel the setup effect is using ratio maps $Z(\vec{r},eV)$ (eq.~\ref{eq:norm_Z})~\cite{mcelroy2003relating, he2014fermi}, which is shown for comparison in Figure~\ref{fig:gNZ_comparison}c.

\begin{equation}
    Z(\vec{r},eV) = g(\vec{r},+E)/g(\vec{r},-E)
\label{eq:norm_Z}
\end{equation}

In this work, we consistently calculate $N(\vec{r},eV)$ before utilizing our denoising algorithm.

\section{Comparison of Noise2Self and Gaussian filtering results for experimental data}\label{app:comparison_N2S_gaus}
\begin{figure}[!h]
    \centering
    \includegraphics[width=0.5\textwidth]{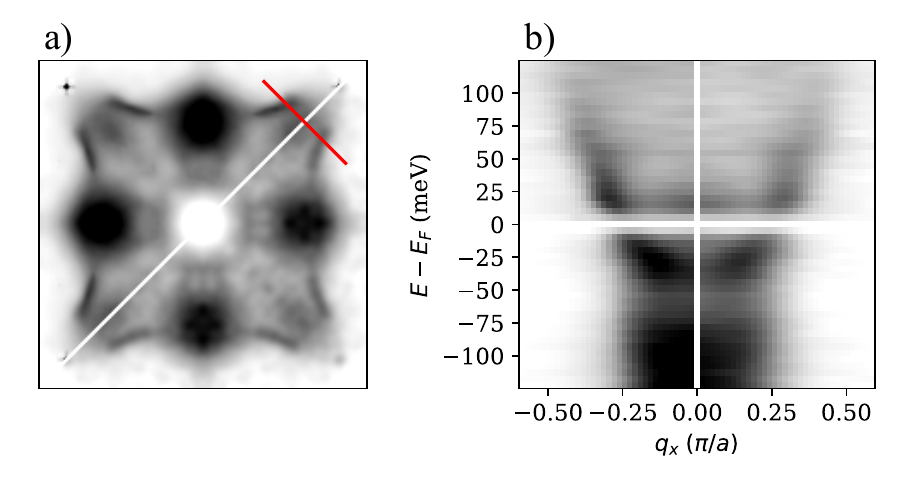}
    \caption{\textbf{Symmetrized results of Noise2Self and Gaussian filtering for the OD3K data.} (a) Comparison of the symmetrized Noise2Self (top left) and Gaussian filtered (bottom right) results for the OD3K Fermi layer. (b) Cut indicated with the red line in (a), with Noise2Self denoised on the left side and Gaussian filtered on the right side.}
    \label{fig:linecuts_gaus}
\end{figure}

In Figure~\ref{fig:linecuts_gaus}, we can make a visual comparison of the denoised results for the experimental OD3K data. We observe that the two methods (Noise2Self and Gaussian filtering) achieve similar results, but the Noise2Self method retains more sharpness and detail (see e.g. the Bragg peaks in Fig.~\ref{fig:linecuts_gaus}a), while the Gaussian filtered image stays slightly noisy. Figure~\ref{fig:linecuts_gaus}b shows a comparison for the antinodal cut indicated with the red line in Figure~\ref{fig:linecuts_gaus}a. We see that the Noise2Self result appears slightly sharper, with more contrast between signal and background.

\bibliography{references}% Produces the bibliography via BibTeX.

\end{document}